\documentclass[openbib,12pt]{article}
\renewcommand{\baselinestretch}{1.5}
\newcommand{\utwi}[1]{\mbox{\boldmath $ #1$}}

\usepackage{booktabs,dcolumn}
\usepackage{psfrag,graphicx}
\usepackage{eso-pic,graphicx}
\usepackage{amsmath}
\usepackage{graphicx}
\usepackage{pdflscape}
\usepackage{longtable}
\usepackage{epstopdf}

\newcolumntype{z}[1]{D{.}{.}{#1}}

\newcommand{\cred}{\textcolor{red}}
\newcommand{\cb}{\textcolor{blue}}
\date{}
\renewcommand{\baselinestretch}{1.5}
\begin{document}

\title{
\begin{center} {\Large \bf Bayesian Realized-GARCH Models for Financial Tail Risk Forecasting Incorporating Two-sided Weibull Distribution } \end{center}}

\author{Chao Wang$^{1}$, Qian Chen$^{2}$, Richard Gerlach$^{1}$  
\\
$^{1}$Discipline of Business Analytics, The University of Sydney\\
$^{2}$HSBC Business School, Peking University} 

\date{} \maketitle

\begin{abstract}
\noindent
The realized GARCH framework is extended to incorporate the two-sided Weibull distribution, for the purpose of volatility and tail risk forecasting in a
financial time series. Further, the realized range, as a competitor for realized variance or daily returns, is employed in the realized GARCH framework.
Further, sub-sampling and scaling methods are applied to both the realized range and realized variance, to help deal with inherent micro-structure
noise and inefficiency. An adaptive Bayesian Markov Chain Monte Carlo method is developed and employed for estimation and forecasting, whose
properties are assessed and compared with maximum likelihood, via a simulation study. Compared to a range of well-known parametric GARCH, GARCH
with two-sided Weibull distribution and realized GARCH models, tail risk forecasting results across 7 market index return series and 2 individual
assets clearly favor the realized GARCH models incorporating two-sided Weibull distribution, especially models employing the sub-sampled realized
variance and sub-sampled realized range, over a six year period that includes the global financial crisis.

\vspace{0.5cm}

\end{abstract}

\newpage
\pagenumbering{arabic}

{\centering
\section{Introduction}
\par
}
\noindent
Since the introduction of Value-at-Risk (VaR) by J.P. Morgan in the RiskMetrics model in 1993, many financial institutions and corporations worldwide
now employ Value-at-Risk (VaR) to assist their decision making on capital allocation and risk management. VaR represents the market risk as one
number and has become a standard risk measurement tool. However, VaR has been criticized, because it cannot measure the expected loss for
extreme, violating returns and is also not mathematically coherent: i.e. it can favour non-diversification. Expected Shortfall (ES), proposed by
Artzner \emph{et al.} (1997, 1999), gives the expected loss, conditional on returns exceeding a VaR threshold, and is a coherent measure. Thus,
in recent years it has become more widely employed for tail risk measurement and is now recommended in the Basel Capital Accord.

Accurate volatility estimation plays a crucial role in parametric VaR and ES calculations. Among the volatility estimation models, the
Autoregressive Conditional Heteroskedasticity model (ARCH) and Generalized ARCH (GARCH) gain high popularity in recent decades, proposed by Engle (1982)
and Bollerslev (1986) respectively. Numerous GARCH-type extension models are also developed during the past few decades: e.g. EGARCH (Nelson, 1991)
and GJR-GARCH (Glosten, Jagannathan and Runkle, 1993) are introduced to capture the well known leverage effect (see e.g. Black, 1976). A second
crucial aspect in parametric tail-risk estimation is the specification of the conditional return, or error, distribution. A voluminous literature shows
this should be heavy-tailed and possibly skewed, see e.g. Bollerslev (1987), Hansen (1994), Chen \emph{et al.} (2012), Chen and Gerlach (2013). Bollerslev (1987) proposed the GARCH with conditional Student-t error distribution. Hansen (1994) developed a skewed Student-t distribution, employing it with a GARCH model, also allowing both
conditional skewness and kurtosis to change over time. Chen \emph{et al.} (2012) employed an asymmetric Laplace distribution (ALD),
combined with a GJR-GARCH model, finding that it was the only consistently conservative tail-risk forecaster, compared with e.g. the Student-t and
Gaussian errors, during the GFC period. Chen and Gerlach (2013) developed a two-sided Weibull distribution, based on the work of Malevergne
and Sornette (2004), which is a natural and more flexible extension of the ALD. They employed this two-sided Weibull as the conditional return
distribution, illustrating its accuracy in tail risk forecasting for index and asset returns, when combined with a range of GARCH-type models.

Various realized measures have been proposed to improve daily volatility estimation, given the wide and increasing availability of high frequency
intra-day data, including Realized Variance (RV): Andersen and Bollerslev (1998), Andersen \emph{et al.} (2003);
and Realized Range (RR): Martens and van Dijk (2007), Christensen and Podolskij (2007). In order to further deal with inherent micro-structure
noise, Martens and van Dijk (2007) and Zhang, Mykland and A\"{i}t-Sahalia (2005) develop methods for scaling and sub-sampling these processes,
respectively, aiming to provide smoother and more efficient realized measures. In Gerlach, Walpole and Wang (2016) the method of sub-sampling is
extended to apply to the realized range.

Hansen \emph{et al.} (2011) extended the parametric GARCH model framework by proposing the Realized-GARCH (Re-GARCH), adding a measurement
equation that contemporaneously links unobserved volatility with a realized measure. Gerlach and Wang (2016) extended the Re-GARCH model
through employing RR as the realized measure and illustrated that the proposed Re-GARCH-RR framework can generate
more accurate and efficient volatility, VaR and ES forecasts compared to traditional GARCH and Re-GARCH models. Watanabe (2012) considered
Student-t and skewed Student-t observation equation errors, whilst Contino and Gerlach (2017) considered those choices also, including
for the measurement equation. Gerlach and Wang (2016) and Contino and Gerlach (2017) found that the Student-t-Gaussian observation-measurement
equation combination, employing the realized range, was the most favoured by predictive likelihood, but was still rejected, in at least half the data
series considered, by standard quantile testing methods.

To improve on this situation, this paper proposes to employ the two-sided Weibull distribution in the
Realized-GARCH framework (Re-GARCH-TWG), motivated by the findings in Chen and Gerlach (2013) and Gerlach and Wang (2016). Further, we
extend the Realized-GARCH modelling framework through incorporating scaled and sub-sampled realized measures, compared with Hansen \emph{et al.} (2011),
Watanabe (2012) and Gerlach and Wang (2016). Further, an adaptive Bayesian MCMC algorithm is developed for the proposed model, extending that in
Gerlach and Wang (2016). To evaluate the performance of the proposed Re-GARCH-TWG model, employing various realized measures as inputs, the accuracy of
the associated VaR and ES forecasts will be assessed and compared with competitors such as the CARE, GARCH and standard Re-GARCH models.

The paper is structured as follows: Section \ref{realized_measure_section} reviews several realized measures and proposes the sub-sampled RR.
A review of the two-sided Weibull distribution, its standardization process and related properties is presented in Section\ref{stw_section}.
Section \ref{model_proposed_section} proposes the Realized-GARCH-TWG model incorporating various realized measures; the associated likelihood and
the adaptive Bayesian MCMC algorithm for estimation and forecasting are presented in Section \ref{parameter_estimation_section}.
The simulation and empirical studies are discussed in Section \ref{simulation_section} and Section \ref{data_empirical_section} respectively.
Section \ref{conclusion_section} concludes the paper and discusses future work.

{\centering
\section{\normalsize REALIZED MEASURES}\label{realized_measure_section}
\par
}
\noindent
This section reviews popular realized measures and also the sub-sampled Realized Range.

For day $t$, representing the daily high, low and closing prices as $H_{t}$, $L_{t}$ and $C_{t}$, the most commonly used daily log return is:
\begin{equation}\label{return_def}
r_t= \text{log}(C_t)-\text{log}(C_{t-1}) \nonumber
\end{equation}
where $r_t^2$ is the associated volatility estimator.


If each day $t$ is divided into $N$ equally sized intervals of length $\Delta$, subscripted by $\Theta= {0, 1, 2, ... , N}$, several
high frequency volatility measures can be calculated. For day $t$, denote the $i$-$th$ interval closing price as $P_{t-1+i \triangle}$ and
$H_{t,i}=\text{sup}_{(i-1) \triangle<j< i \triangle}P_{t-1+j}$ and $L_{t,i}=\text{inf}_{(i-1) \triangle<j<i \triangle}{P_{t-1+j}}$ as the high and
low prices during this time interval. Then RV, as proposed by Andersen and Bollerslev (1998) is then:
\begin{equation}\label{rv_def2}
RV_{t}^{\triangle}=\sum_{i=1}^{N} [log(P_{t-1+i \triangle})-log(P_{t-1+(i-1)\triangle})]^{2}
\end{equation}
Martens and van Dijk (2007) and Christensen and Podolskij (2007) developed the Realized Range, which sums the squared intra-period
ranges:
\begin{equation}\label{rrv_def}
RR_{t}^{\triangle}= \frac {\sum_{i=1}^{N}(\text{log}H_{t,i}-\text{log}L_{t,i})^2}{4\log2} \, .
\end{equation}
Through theoretical derivation and simulation, Martijns and van Dijk (2007) show that RR is a competitive, and sometimes more efficient,
volatility estimator than RV, under some micro-structure conditions and levels. Gerlach and Wang (2016) confirm that RR can provide increased predictive
likelihood performance, and improved accuracy and efficiency in empirical tail risk forecasting, when employed as the measurement equation variable
in an Re-GARCH model.

To further reduce the effect of microstructure noise, Martens and van Dijk (2007) presented a scaling process, as in Equations (\ref{rv_scale})
and (\ref{rrv_scale}).
\begin{eqnarray}\label{rv_scale}
RV_{S,t}^{\triangle}= \frac {\sum_{l=1}^{q}RV_{t-l}}{\sum_{l=1}^{q}RV_{t-l}^{\triangle}}RV_{t}^{\triangle},
\end{eqnarray}
\begin{eqnarray}\label{rrv_scale}
RR_{S,t}^{\triangle}= \frac {\sum_{l=1}^{q}RR_{t-l}}{\sum_{l=1}^{q}RR_{t-l}^{\triangle}}RR_{t}^{\triangle},
\end{eqnarray}
\noindent
where $RV_{t}$ and $RR_{t}$ represent the daily squared return and squared range on day $t$, respectively, and $q$ is selected as 66. This scaling
process is inspired by the fact that the daily squared return and range are each less affected by micro-structure noise than their high
frequency counterparts, thus can be used to scale and smooth RV and RR, creating less micro-structure sensitive measures.

Further, Zhang, Mykland and A\"{i}t-Sahalia (2005) proposed a sub-sampling process, also to deal with micro-structure effects.
For day $t$, $N$ equally sized samples are grouped into $M$ non-overlapping subsets $\Theta^{(m)}$ with size $N/M=n_{k}$, which means:
\begin{equation}
\Theta= \bigcup_{m=1}^{M} \Theta^{(m)}, \; \text{where} \; \Theta^{(k)}  \cap \Theta^{(l)} = \emptyset, \;
\text{when}  \;  k \neq l.  \nonumber
\end{equation}
Then sub-sampling will be implemented on the subsets $\Theta^{i}$ with $n_{k}$ interval:
\begin{equation}
\Theta^{i}= {i, i+n_k,...,i+n_k(M-2), i+n_k(M-1)}, \; \text{where} \;  i= {0,1,2...,n_k-1}.  \nonumber
\end{equation}

Representing the log closing price at the $i$-$th$ interval of day $t$ as $C_{t,i}=P_{t-1+i\triangle}$, the RV with the subsets
$\Theta^{i}$ is:
\begin{equation}
RV_{i}= \sum_{m=1}^{M} (C_{t,i+n_{k}m}-C_{t,i+n_{k}(m-1)})^{2}; \; \text{where} \; i= {0,1,2...,n_k-1}.  \nonumber
\end{equation}

We have the $T/M$ RV with $T/N$ sub-sampling as (supposing there are $T$ minutes per trading day):

\begin{equation}
RV_{T/M,T/N}= \frac{\sum_{i=0}^{n_k-1} RV_{i} } {n_k},
\end{equation}

Then, denoting the high and low prices during the interval $i+n_{k}(m-1)$ and $i+n_{k}m$ as
$H_{t,i}=\text{sup}_{(i+n_{k}(m-1))\triangle<j<(i+n_{k}m) \triangle}P_{t-1+j}$ and
$L_{t,i}=\text{inf}_{(i+n_{k}(m-1))\triangle<j<(i+n_{k}m) \triangle}{P_{t-1+j}}$ respectively, we propose the $T/M$ RR with $T/N$ sub-sampling
as:
\begin{equation}
RR_{i}= \sum_{m=1}^{M} (H_{t,i}-L_{t,i})^2; \; \text{where} \; i= {0,1,2...,n_k-1}.
\end{equation}
\begin{equation}
RR_{T/M,T/N}= \frac{\sum_{i=0}^{n_k-1} RR_{i} } {4 \text{log}2 n_k},
\end{equation}

For example, the 5 mins RV and RR with 1 min subsampling can be calculated as below respectively:
\begin{eqnarray}  \nonumber
&&RV_{5,1,0}=(\text{log} C_{t5}-\text{log} C_{t0})^2+(\text{log} C_{t10}-\text{log} C_{t5})^2+... \\  \nonumber
&&RV_{5,1,1}=(\text{log}C_{t6}-\text{log} C_{t1})^2+(\text{log} C_{t11}-\text{log} C_{t6})^2+... \\ \nonumber
&&RV_{5,1}=\frac{\sum_{i=0}^{4}RV_{5,1,i}} {5}  \nonumber
\end{eqnarray}
\begin{eqnarray}  \nonumber
&&RR_{5,1,0}=(\text{log}H_{t0\leq t \leq t5}-\text{log} L_{t0\leq t \leq  t5})^2+(\text{log}H_{t5\leq t \leq t10}-\text{log}
L_{t5\leq t \leq  t10})^2+... \\ \nonumber
&&RR_{5,1,1}=(\text{log} H_{t1\leq t \leq t6}-\text{log} L_{t1\leq t \leq  t6})^2+(\text{log}H_{t6\leq t \leq t11}-\text{log}
L_{t6\leq t \leq  t11})^2+... \\ \nonumber
&&RR_{5,1}=\frac{\sum_{i=0}^{4}RR_{5,1,i}} {4 \text{log} (2)5} \nonumber
\end{eqnarray}

{\centering
\section{A two-sided Weibull distribution}\label{stw_section}
\par
}
\noindent
The Weibull distribution, introduced by Weibull (1951), is a special case of an extreme value distribution
and of the generalized gamma distribution. It is widely applied in the fields of material science, engineering and also in
finance, due to its versatility. Mittnik and Ratchev (1989) found it to be the most accurate for the unconditional
return distribution for the S\&P500 index when applied separately to positive and negative returns; while various authors
have employed it as an error distribution in range data modelling (see Chen \emph{et al.}, 2008) and autoregressive conditional duration (ACD) models
(see e.g. Engle and Russell, 1998).

The TW's shape and scale is tuned by four parameters. The general definition of a TW distribution is,
$Y \sim TW(\lambda_1,k_1,\lambda_2,k_2)$ if:
\begin{eqnarray*}
  \left\{ \begin{array}{ll}
  -Y  \sim     \mbox{Weibull} (\lambda_1,k_1)    \,\,;\,\, Y<0 \\
   Y  \sim     \mbox{Weibull} (\lambda_2,k_2)    \,\,;\,\, Y \ge 0 \end{array} \right.
\end{eqnarray*}
Here the shape parameters satisfy $k_1, k_2 > 0$ and scale parameters $\lambda_1, \lambda_2 > 0$.

\subsection{Standardized Two-sided Weibull distribution}\label{tw_standardisation}
Since the observation error in a GARCH-type model needs to have mean 0 and variance 1, Chen and Gerlach (2013)
developed the standardized two-sided Weibull distribution (STW), subsequently deriving the pdf, cdf, quantile function
and the conditional expectation functions required to calculate the likelihood, VaR and ES measures for the STW distribution.

A standardized TW distribution is equivalent to $X = \frac{Y}{\sqrt{\mbox{Var}(Y)}}$, where $Y \sim TW(\lambda_1,k_1,\lambda_2,k_2)$.
It can be shown that:
$$
\mbox{Var}(Y) = b_p^2 = \frac{\lambda^3_1}{k_1}\Gamma\left(1+\frac{2}{k_1}\right)+\frac{\lambda^3_2}{k_2}\Gamma\left(1+\frac{2}{k_2}\right)-
\left[-\frac{\lambda^2_1}{k_1}\Gamma\left(1+\frac{1}{k_1}\right)+\frac{\lambda^2_2}{k_2}\Gamma\left(1+\frac{1}{k_2}\right)\right]^2.
$$
The pdf for an STW random variable $X=\frac{Y}{\sqrt{\mbox{Var}(Y)}}$, where $Y \sim TW(\lambda_1,k_1,\lambda_2,k_2)$, is:
\begin{eqnarray}\label{StdTWpdf}
f(x|\lambda_1,k_1,k_2)=\left\{ \begin{array}{ll}
b_p\left(\frac{-b_p x}{\lambda_1}\right)^{k_1-1}\exp\left[-\left(\frac{-b_p x}{\lambda_1}\right)^{k_1}\right]\,\,;\,\, x<0 \\
b_p\left(\frac{b_p x}{\lambda_2}\right)^{k_2-1}\exp\left[-\left(\frac{b_p x}{\lambda_2}\right)^{k_2}\right]\,\,;\,\, x \ge 0 \end{array} \right.
\end{eqnarray}
To ensure the pdf integrates to 1:
\begin{equation}\label{rest}
\frac{\lambda_1}{k_1}+\frac{\lambda_2}{k_2}=1
\end{equation}

Chen and Gerlach (2013) set $k_1=k_2$ for parsimony and simplification: the choice was well supported by the data; the same specification is made here.
Thus, based on Equation (\ref{rest}), we denote an $STW(\lambda_1,k_1)$, with only
two parameters to estimate. As $Pr(X<0)=\frac{\lambda_1}{k_1}$, thus $0 < \lambda_1 \le  k_1$, and
$\lambda_2=k_1-\lambda_1$.

The mean of an STW, $\mu_X= \frac{-\lambda^2_1}{b_pk_1}\Gamma\left(1+\frac{1}{k_1}\right)+\frac{\lambda^2_2}{b_pk_1}\Gamma\left(1+\frac{1}{k_1}\right)$.
Thus $Z=X-\mu_X$ has a shifted $STW(\lambda_1,k_1)$ distribution with mean 0 and variance 1. The CDF, and other relevant
characteristics of the STW distribution such as skewness and kurtosis, can be found in Chen and Gerlach (2013).

Through employing the STW in the GARCH framework, Chen and Gerlach (2013) found that their proposed models perform at least as well
as other distributions for VaR forecasting, but perform most favourably for expected shortfall forecasting, prior to, as well as during and
after, the 2008 global financial crisis.

{\centering
\section{Model Proposed}\label{model_proposed_section}
\par
}
\noindent
The realized GARCH model of Hansen \emph{et al.} (2011) can be written as:
\begin{eqnarray}\label{rgarch}
&& r_t= \sqrt{h_t} z_t, \\ \nonumber
&& h_t= \omega +\beta h_{t-1}+ \gamma x_{t-1} \,\,  , \\ \nonumber
&& x_t = \xi +\varphi h_{t}+ \tau_1 z_t + \tau_2 (z_t^2-1)+  \sigma_{\varepsilon} \varepsilon_t \,\,  ,
\end{eqnarray}
where $r_t= [\text{log}(C_t)-\text{log}(C_{t-1})]\times100$ is the percentage log-return for day $t$,
$z_t \stackrel{\rm i.i.d.} {\sim} D_1(0,1)$ and $ \varepsilon_t \stackrel{\rm i.i.d.} {\sim} D_2(0,1)$ and
$x_t$ is a realized measure, e.g. RV; $D_1(0,1), D_2(0,1)$ indicate distributions that have mean 0 and variance 1.
The three equations in order in model (\ref{rgarch}) are: the \emph{return equation},
the \emph{volatility equation} and the \emph{measurement equation}, respectively. The measurement equation is a second observation
equation that captures the contemporaneous dependence between latent volatility and the realized measure.
The term $\tau_1 z_t + \tau_2 (z_t^2-1)$ is used to capture the leverage effect.

Hansen \emph{et al.} (2011) utilize the RV as the realized measure $x_t$ in model (\ref{rgarch});
and chose Gaussian errors, e.g. $D_1(0,1) = D_2(0,1) \equiv N(0,1)$. Watanabe (2012) allowed $D_1(0,1)$ to be a standardised
Student-t; Contino and Gerlach (2017) allowed it to be the skewed-t of Hansen (1994) and also allowed $D_2(0,1)$ to be a
standardised Student-t. Gerlach and Wang (2016) proposed Realized Range Re-GARCH (RR-RG) via the choice of $x_t = RR_{t}^{\triangle}$.
The choice of RR as information to drive volatility is motivated by Martijns and van Dijk (2007).

In this paper, we extend the realized GARCH model through incorporating $D_1(0,1)$ as an STW distribution. This proposed class of models is
subsequently denoted as Re-GARCH-TWG or RG-TWG. Further, the scaled and sub-sampled realized measures, as presented in
Section \ref{realized_measure_section}, are employed as $x_t$ in the realized GARCH framework in this paper.

Stationarity is an important issue in time series modelling. As derived in Hansen \emph{et al.} (2011) and Gerlach and Wang (2016), the
required stationarity conditions for the general realized GARCH model are:
\begin{eqnarray} \label{stationary_condition}
&& \omega+ \gamma \xi>0, \\  \nonumber
&& 0<\beta+\gamma \varphi<1
\end{eqnarray}
\noindent

To ensure positivity of each $h_t$, it is sufficient that $\omega, \beta, \gamma$ are all positive. Further, as discussed in
Section \ref{tw_standardisation}, the constraint $0 < \lambda_1 \le  k_1$ is also incorporated. This set of conditions is subsequently enforced
during estimation of all proposed realized GARCH models employing STW distribution in this paper.

{\centering
\section{\normalsize LIKELIHOOD AND BAYESIAN ESTIMATION} \label{parameter_estimation_section}
\par
}
\noindent

\subsection{Likelihood}
Following Hansen \emph{et al.} (2011), where $D_1 = D_2 \equiv N(0,1)$, the log-likelihood function for model (\ref{rgarch}) is:
\begin{equation}\label{RGGG_lik}
\ell (r,x;\theta)=\underbrace{ -\frac {1}{2} \sum_{t=1}^{n} \left[ log(2 \pi)+log(h_t)+ r_t^2 / h_t \right]}_{\ell (r;\theta)}
 \underbrace{-\frac {1}{2} \sum_{t=1}^{n} \left[ log(2 \pi)+log(\sigma_{\varepsilon}^2)+ \varepsilon_t^2/\sigma_{\varepsilon}^2
 \right]}_{\ell (x|r;\theta)}
\end{equation}
where $\varepsilon_t=x_t-\xi -\varphi h_{t}-\tau_1 z_t - \tau_2 (z_t^2-1)$; the parameter vector to be estimated
is $\theta=(\omega,\beta,\gamma,\xi,\varphi,\tau_1,\tau_2,\sigma_{\varepsilon})^{'}$. Hansen \emph{et al.} (2011) derived the 1st and
2nd derivative of this log-likelihood function, allowing calculation of asymptotic standard errors of estimation, via a Hessian matrix.
Subsequently, this model is denoted RG-GG (Realized GARCH with Gaussian-Gaussian errors).

Under our choice $D_1 \sim STW(0,1); \, D_2 \equiv N(0,1)$, as in Equation (\ref{StdTWpdf}), the log-likelihood
function for model (\ref{rgarch}) is:
\begin{eqnarray}\label{RGTWG_lik_1}
\ell (r,x;\theta) &=& \underbrace{ n log(b_p) + \sum_{t=1}^{n} (k_1-1) log\left(\frac{-b_p x}{\lambda_1} \right)- \sum_{t=1}^{n} \left( \frac{-b_p x}{\lambda_1} \right)^{k_1} }_{\ell (r;\theta)}  \\
  && \underbrace{-\frac {1}{2} \sum_{t=1}^{n} \left[ log(2 \pi)+log(\sigma_{\varepsilon}^2)+ \varepsilon_t^2/\sigma_{\varepsilon}^2
 \right]}_{\ell (x|r;\theta)}; \nonumber
\end{eqnarray}
when $x<0$, and

\begin{eqnarray}\label{RGTWG_lik_2}
\ell (r,x;\theta) &=& \underbrace{ n log(b_p) + \sum_{t=1}^{n} (k_1-1) log\left(\frac{b_p x}{\lambda_2} \right)- \sum_{t=1}^{n} \left( \frac{b_p x}{\lambda_2} \right)^{k_1} }_{\ell (r;\theta)}  \\
  && \underbrace{-\frac {1}{2} \sum_{t=1}^{n} \left[ log(2 \pi)+log(\sigma_{\varepsilon}^2)+ \varepsilon_t^2/\sigma_{\varepsilon}^2
 \right]}_{\ell (x|r;\theta)}; \nonumber
\end{eqnarray}
when $x\ge 0$. Here $b_p^2= Var(Y)$, $\lambda_2=k_1-\lambda_1$. $X = Y / b_p$ is standardised, and
$\varepsilon_t=x_t-\xi -\varphi h_{t}-\tau_1 z_t - \tau_2 (z_t^2-1)$.

The parameter vector to be estimated is now $\utwi{\theta}=(\omega,\beta,\gamma, \lambda_1, k_1, \xi,\varphi,\tau_1,\tau_2,\sigma_{\varepsilon})^{'}$,
under the constraints in (\ref{stationary_condition}) and positivity on $(\omega, \beta, \gamma)$; further we restrict $0 < \lambda_1 \le  k_1$ .

\subsection{Bayesian estimation methods}
\noindent
This section specifies the Bayesian methods and MCMC procedures for estimating parameters.
The likelihoods in (\ref{RGTWG_lik_1}) and (\ref{RGTWG_lik_2}) involve 10 unknown parameters; most of which are
part of equations involving latent, unobserved variables. The performance and finite sample properties of ML estimates of
these likelihoods are not yet studied. As such, we also consider powerful numerical and computational algorithms in a Bayesian
framework, under uninformative priors, as a competing estimator for these models.

An adaptive MCMC method, extended and adapted from that in Gerlach and Wang (2016), is employed, based on the
"epoch" method in Chen \emph{et al.} (2017). For the initial "epoch" of the burn-in period, a Metropolis algorithm (Metropolis \emph{et al.}, 1953) employing a
mixture of 3 Gaussian proposal distributions, with a random walk mean vector, is utilised for each block of parameters. The proposal var-cov matrix of each block in each mixture element is $C_i \Sigma$, where $C_1 =1;C_2 =100;C_3 =0.01$, with $\Sigma$ initially set to $\frac{2.38}{\sqrt{(d_i)}}I_{d_i}$, where $d_i$ is the dimension of the block ($i$) of parameters being generated, and
$I_{d_i}$ is the identity matrix of dimension $d_i$. This covariance matrix is subsequently tuned, aiming towards a
target acceptance rate of $23.4\%$ (if $d_i>4$, or $35\%$ if $2 \le d_i \le 4$, or $44\%$ if $d_i=1$), as standard, via the algorithm of
Roberts, Gelman and Gilks (1997).

In order to enhance the convergence of the chain, at the end of 1st epoch (say 20,000 iterations), the covariance matrix for each parameter block
is calculated, after discarding (say) the first 2,000 iterations, which is used in the proposal distribution in the next epoch (of 20,000 iterations).
After each epoch, the standard deviations of each parameter chain in that epoch are calculated and compared to those form the previous epoch. This
process is continued until the mean absolute percentage change is less than a pre-specified threshold, e.g. 10\%.
In the empirical study, on average it takes 3-4 Epochs to observe this absolute percentage change lower than 10\%; thus, the chains are run in
total for 60,000-80,000 iterations as a burn-in period, in the empirical parts of this paper. A final epoch is run, of say 10,000 iterates,
employing a mixture of three Gaussian proposal distributions, in an "independent" Metropolis-Hastings algorithm, in each block. The mean vector
for each block is set as the sample mean vector of the last epoch iterates (after discarding the first 2,000 iterates) for that block; i.e. it is the same for each of the
three mixture elements. The proposal var-cov matrix in each element is $C_i \Sigma$, where $C_1 =1;C_2 =100;C_3 =0.01$ and $\Sigma$ is the sample
covariance matrix of the last epoch iterates for that block (after discarding the first 2,000 iterates).

As an example, for the RG-TWG model, three blocks were employed: $\utwi{\theta_1}=(\omega,\beta,\gamma, \varphi)^{'}$,
$\utwi{\theta_2}=(\xi,\tau_1,\tau_2,\sigma)^{'}$ and $\utwi{\theta_3}=(\lambda_1,k_1)$, via the motivation that parameters within the
same block are more strongly correlated in the posterior (likelihood) than those between blocks: e.g. the
stationarity condition causes correlation between iterates of $\beta,\gamma,\varphi$, thus they are kept together in a block.

Priors are chosen to be uninformative over the possible stationarity and positivity regions,
e.g. $\pi(\utwi{\theta})\propto I(A)$, which is a flat prior for $\utwi{\theta}$ over the region $A$.

{\centering
\section{Simulation study}\label{simulation_section}
\par
}
\noindent
A simulation study is presented to illustrate the comparative performance of the MCMC and ML estimators,
in terms of parameter estimation, quantile and expected shortfall forecasting accuracy. The aim is to
illustrate the bias and precision properties for these two methods, highlighting the comparative
performance of the MCMC estimator.

5000 replicated data sets of size $n=3000$ are simulated from the following specific RG-TWG model:
\begin{eqnarray*}
\mbox{\textbf Model 1}&& r_t= \sqrt{h_t} z_t, \,\, z_t \sim STW(0.6,1.1) \\
&& h_t= 0.02 + 0.25 h_{t-1}+ 0.75 x_{t-1} \,\,  , \\
&& x_t = 0.1 + 0.95 h_{t}+ 0.1 z_t -0.1 (z_t^2-1)+  \varepsilon_t \\
&& \varepsilon_t \sim N(0,0.5^2) \, .
\end{eqnarray*}
In Model 1 $r_{t}$ is analogous to a daily log-return and $x_t$ is analogous to the daily realized measure. The persistence
level ($\beta+\gamma \varphi$) is deliberately chosen very close to 1; with all true values close to those
estimated from real data. For each model the forecast $\alpha$-level VaR and ES detail is presented in Chen and Gerlach (2013).
Following Basel II and Basel III risk management guidelines, the quantile level $\alpha=0.01$ is considered. For both estimation methods,
all initial parameter values were arbitrarily set equal to $0.25$.

Estimation results are summarised in Table \ref{simu_stat1}. Boxes indicate the optimal measure
comparing MCMC and ML for both bias (Mean) and precision (RMSE). The results are clearly in favour of the MCMC method overall.
The bias results favoured MCMC with 9 out of 10 parameter estimates and one-step-ahead ES forecasting; whilst the MCMC method precision is
also much higher for all 10 parameters and one-step-ahead VaR \& ES forecasts. This highlights convergence, bias and precision issues with the MLE
that are greatly improved via the MCMC approach, which is employed afterwards in the empirical study.

\renewcommand{\baselinestretch}{1.1}
\begin{table}[!ht]
\begin{center}
\caption{\label{simu_stat1} \small Summary statistics for the two estimators of the RG-TWG model, data simulated from Model 1.}\tabcolsep=10pt
\begin{tabular}{lcccccc} \hline
$n=5000$    &        &\multicolumn{2}{c}{MCMC}     &  \multicolumn{2}{c}{ML}   \\
Parameter   &  True  &      Mean    &       RMSE   &        Mean   &      RMSE    \\ \hline
$\omega$    &  0.02	 &\fbox{0.1150} &\fbox{0.1302} &	   0.1514  &  0.5288  \\
$\beta$     &  0.75	 &\fbox{0.7457} &\fbox{0.0146} &	   0.7286  &  0.0896   \\
$\gamma$    &  0.25	 &\fbox{0.2275} &\fbox{0.0387} &	   0.4149  &  0.7986 \\
$\xi$       &  0.10	 &     -0.3414  &\fbox{0.6840} &\fbox{-0.2330} &  1.0298 \\
$\varphi$   &  0.95	 &      1.0764  &\fbox{0.2042} & \fbox{0.9933} &  0.3429 \\
$\tau_1$    & -0.02	 &\fbox{-0.0200}&\fbox{0.0098} & 	  -0.0212  &  0.0218  \\
$\tau_2$    &  0.02	 &\fbox{0.0201} &\fbox{0.0049} &       0.0344  &  0.1497 \\
$\sigma$    &  0.50	 &\fbox{0.5047} &\fbox{0.0082} &	   0.5126  &  0.1298 \\
$\lambda_1$ &  0.60	 &\fbox{0.6003} &\fbox{0.0288} &	   0.5753  &  0.0793 \\
$k_1$       &  1.10	 &\fbox{1.1004} &\fbox{0.0260} &	   1.0777  &  0.0908 \\
$\text{VaR}_{t+1}$ & -4.9442 &      -4.9744 &\fbox{0.1990}&\fbox{-4.9666}&	1.7084  \\
$\text{ES}_{t+1}$  & -6.1001 &\fbox{-6.1374}&\fbox{0.2553}&	     -6.1402 &	2.6486  \\
 \hline
\end{tabular}
\end{center}
\emph{Note}:\small  A box indicates the favored estimators, based on mean and RMSE.
\end{table}

{\centering
\section{Empirical study}\label{data_empirical_section}
\par
}
\noindent
\subsection{Data}
\noindent
Daily and high frequency data, observed at 1-minute and 5-minute frequency, including daily open, high, low and closing prices, are downloaded from
Thomson Reuters Tick History. Data are collected for 7 market indices: S\&P500, NASDAQ (both US), Hang Seng (Hong Kong), FTSE 100 (UK),
DAX (Germany), SMI (Swiss) and ASX200 (Australia), with time range Jan 2000 to June 2016; as well as for 2 individual assets: IBM and GE (both US).
IBM has the same starting data as 7 indices, while the starting data collection time for GE is May 2000, only after its $3:1$ stock split in May, 2000.

\begin{figure}[htp]
     \centering
\includegraphics[width=.9\textwidth]{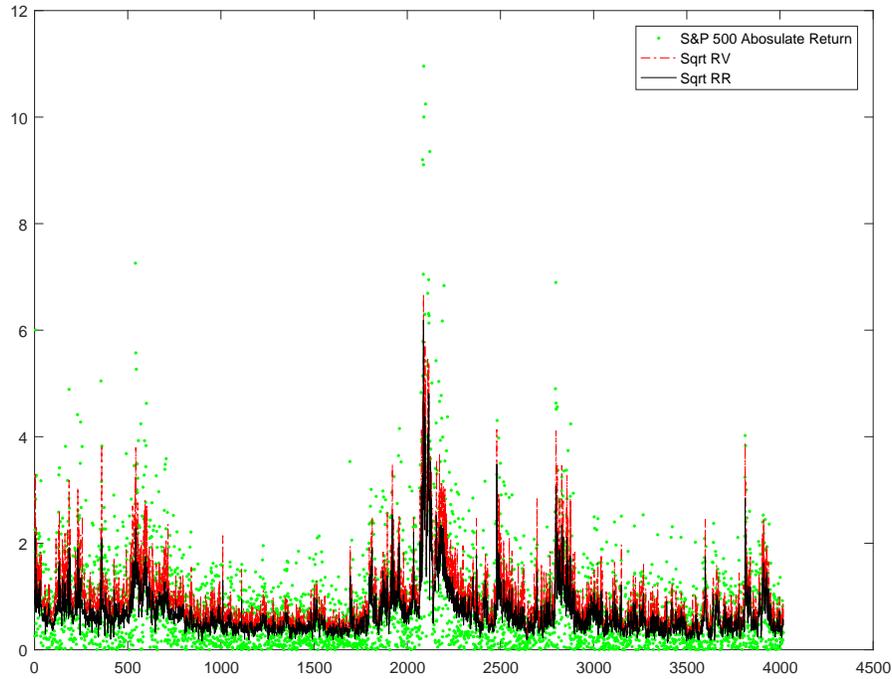}
\caption{\label{Fig2} S\&P 500 absolute return, $\sqrt{RV}$ and $\sqrt{RR}$ Plots.}
\end{figure}

The data are used to calculate the daily return. Further, the 5-minute data
are employed to calculate the daily RV and RR measures, while both 5 and 1-minute data are employed to produce daily scaled and sub-sampled
versions of these two measures, as in Section \ref{realized_measure_section}; $q=66$ is employed for the scaling process, i.e. around 3 months.
Thus, the final starting time is 3 months from the starting time of data collection. Figure \ref{Fig2} plots the S\&P 500 absolute value of daily
returns, as well as $\sqrt{RV}$ and $\sqrt{RR}$, for exposition.

\subsection{Tail Risk Forecasting and Capital Efficiency}
\noindent
After the GFC, with decreased investors' confidence and slowdown of global economic growth, there is less flow of cheap capital; meanwhile, the
stricter regulation (Basel III, fully effective in 2019) puts more pressure on the accuracy and usage of regulatory capital. While regulatory capital
is usually advised by the regulators, the institutions may have internal capital-adequacy-assessment to determine the economic capital for daily
business decision making. Thus, economic capital allocation is usually more dynamic, and could be 90\% to 120\% of regulatory capital according
to a capital-management survey on more than 25 European banks conducted by McKinsey in 2012. The financial institutions can
quickly raise substantial funds for investment by eliminating excess conservatism through accurate calculation of risk levels. We aim to show that
our model can provide these.

The Basel II and III Capital Accords favour VaR and ES as tail risk measures for financial institutions to employ in market risk management.
Therefore, it is very important for institutions to have access to highly accurate VaR and ES forecast models, allowing accurate
capital allocation, both to avoid default and over-allocation of funds. Both daily Value-at-Risk (VaR) and Expected Shortfall (ES) are estimated
for the 7 indices and the 2 asset series, as recommended in the Basel II and III Capital Accord.

The $1$-period VaR, for holding an asset, and the conditional $1$-period VaR, or ES, are formally defined via
\begin{eqnarray*}
\alpha = Pr( r_{t+1} < \mbox{VaR}_{\alpha} | \Omega_{t} )  \,\,\,;\,\, \mbox{ES}_{\alpha} = E\left[ r_{t+1} | r_{t+1} < \mbox{VaR}_{\alpha},\Omega_{t} \right]
\end{eqnarray*}
where $r_{t+1}$ is the one-period return from time $t$ to time $t+1$, $\alpha$ is the quantile level and $\Omega_{t}$ is the
information set at time $t$.

A rolling window with fixed size in-sample data is employed for estimation to produce each 1 step ahead forecast; the in-sample size $n$ is
given in Table \ref{var_fore_table} for each series, which differs due to non-trading days in each market. In order to see the
performance during the GFC period, the initial date of the forecast sample is chosen as the beginning of 2008. On average,
2111 VaR and ES forecasts are generated for each return series from a range of models. These include the proposed Re-GARCH-TWG type models
(estimated with MCMC) with different input measures of volatility: RV \& RR, scaled RV \& RR and
sub-sampled RV \& RR. The conventional GARCH, EGARCH and GJR-GARCH with
Student-t distribution, CARE-SAV (Taylor, 2008) and Re-GARCH with RV and Gaussian or Student-t observation error distributions,
are also included, for the purpose of comparison. Further, a filtered GARCH (GARCH-HS) approach is also included,
where a GARCH-t is fit to the in-sample data, then a standardised VaR and ES are estimated via historical simulation
from the sample of returns (e.g. $r_1,\ldots,r_n$) divided by their GARCH-estimated conditional standard deviation (i.e. $r_t/\sqrt{\hat{h_t}}$).
Then final forecasts of VaR, ES are found by multiplying the standardised VaR, ES estimates by the forecast $\sqrt{\hat{h}_{n+1}}$ from the GARCH-t model.
All these aforementioned models are estimated by ML, using the Econometrics toolbox in Matlab (GARCH-t, EGARCH-t, GJR-t and GARCH-HS) or code
developed by the authors (CARE-SAV and Re-GARCH). The GARCH-TW model of Chen and Gerlach (2013) is also included, estimated by the
MCMC scheme proposed in this paper. The actual forecast sample sizes $m$, in each series, are given in Table \ref{var_fore_table}.

The VaR violation rate (VRate) is employed to initially assess the VaR forecasting accuracy. VRate is simply the proportion of returns
that exceed the forecasted VaR level in the forecasting period, given in Equation (\ref{varvrate_equation}):
models with VRate closest to nominal quantile level $\alpha=0.01$ are preferred. Regarding models with VRate with the same absolute
distances to 1\% nominal quantile level, the one that is conservative is preferred, e.g 0.95\% is preferred compared with 1.05\%.

\begin{equation}\label{varvrate_equation}
\text{VRate}= \frac{1}{m} \sum_{t=n+1}^{n+m} I(r_t<\text{VaR}_t)\, ,
\end{equation}

\begin{equation}\label{varvrate_equation}
\text{ESRate}= \frac{1}{m} \sum_{t=n+1}^{n+m} I(r_t<\text{ES}_t)\, ,
\end{equation}
where $n$ is the in-sample size and $m$ is the forecasting sample size.

Several standard quantile accuracy and independence tests are also employed: e.g. the unconditional coverage (UC) and conditional
coverage (CC) tests of Kupiec (1995) and Christoffersen (1998) respectively; the dynamic quantile (DQ) test of
Engle and Manganelli (2004); and the Value-at-Risk quantile regression (VQR) test of Gaglione \emph{et al.} (2011).

\subsection{VaR and ES for two-sided Weibull}

The inverse cdf or quantile function (VaR) of an STW is given in (\ref{STWinvcdf}) for the STW distribution:
\begin{eqnarray}\label{STWinvcdf}
F^{-1}(\alpha|\lambda_1,k_1,k_2)=\left\{ \begin{array}{ll}
-\frac{\lambda_1}{b_p}\left[-\ln\left(\frac{k_1}{\lambda_1}\alpha\right)\right]^{\frac{1}{k_1}}\,\,;\,\, 0\le \alpha < \frac{\lambda_1}{k_1} \\
\frac{\lambda_2}{b_p}\left[-\ln\left(\frac{k_1}{\lambda_2}\left(1-\alpha\right)\right)\right]^{\frac{1}{k_1}}\,\,;\,\, \frac{\lambda_1}{k_1} \le \alpha < 1 \end{array} \right.
\end{eqnarray}

In practice, returns are only mildly skewed, therefore, the estimated values for $\frac{\lambda_1}{k_1}$ are close to 0.5.
Since risk management focuses on the extreme tails, e.g. $\alpha \le 0.05$ for a long position, thus only the case
$\alpha < \frac{\lambda_1}{k_1}$ in (\ref{STWinvcdf}) is relevant here. In this case, the ES of the STW is:
\begin{eqnarray*}
ES_{\alpha}=\int^{VaR_{\alpha}}_{-\infty}xf\left(x|x < VaR_{\alpha}\right)dx ,
\end{eqnarray*}
where $f\left(x|x < VaR_{\alpha}\right)$ is the conditional density function, which becomes:
\begin{eqnarray}\label{StTSWeibullES}
\mbox{ES}_{\alpha} &=& \frac{-\lambda^2_1}{\alpha b_pk_1}\int^{\infty}_{\left(\frac{-b_pVaR_{\alpha}}{\lambda_1}\right)^{k_1}}\left[\left(\frac{-b_px}{\lambda_1}\right)^{k_1}\right]^{\frac{1}{k_1}+1-1}\exp\left[-\left(\frac{-b_px}{\lambda_1}\right)^{k_1}\right]d\left(\frac{-b_px}{\lambda_1}\right)^{k_1}\nonumber \\
&=& \frac{-\lambda^2_1}{\alpha b_pk_1}\Gamma\left(1+\frac{1}{k_1},\left(\frac{-b_pVaR_{\alpha}}{\lambda_1}\right)^{k_1}\right)\,\,;\,\, 0 \le \alpha < \frac{\lambda_1}{k_1}
\end{eqnarray}
where $\Gamma(s,x)=\int^{\infty}_{x}t^{s-1}e^{-t}dt$ is the upper incomplete gamma function. The derivation details can be found in Chen and
Gerlach (2013).

\subsubsection{\normalsize Value at Risk}
\noindent
Table \ref{var_fore_table} presents the VRates at the 1\% quantile for each model over the 9 return series,
while Table \ref{Summ_var_fore} summarizes those results. A box indicates the model that has observed VRate closest to 1\% in
each market, while bolding indicates the model with VRate furthest from 1\%. The G-t, EGARCH-t, GJR-t, CARE-SAV, Re-GARCH-GG and
Re-GARCH-tG with RV are estimated with ML, and the Re-GARCH-TWG type models are estimated with MCMC as discussed in
Section \ref{parameter_estimation_section}.

Clearly from Table \ref{var_fore_table}, Re-GARCH-TWG models as a group have most of the optimal VaR forecast series and consistently conservative,
in terms of being closest to VRate of 1\%, over the 9 return series. Based on Table \ref{Summ_var_fore}, Re-GARCH-TWG models employing RV has the
second best VRate (0.869\%) on average and via the median (0.804\%), and GARCH employing the STW distribution (Chen and Gerlach, 2013) has average
VRate 1\% which is closest to 1\%. Later, we will compare Re-GARCH-TWG and G-TW in details, and provide evidence on why Re-GARCH-TWG type model is
preferred in VaR forecasting. The VaR violation rates for different models were further split by index and asset in order to see which models are
preferred by index and asset as in \ref{var_fore_table}, and similar results are observed.

All models, besides RG-TWG and G-TW, on this measure are anti-conservative, having VRates on average (and median) above 1\%: Re-GARCH-GG was
most anti-conservative, generating 80-90\% more violations than exppected, not surprising since it is the only model
employing a Gaussian observation error distribution.

Chang et al. (2011) and McAleer et al. (2013) proposed using forecast combinations of the VaR series from different models, to take advantage
of associated empirically-observed efficiencies from forecast combination, but also to potentially robustify against the effects of
financial crises like the GFC. This approach is employed here: specifically, the four series created by taking the mean ("FC-Mean"),
median ("FC-Med"), minimum ("FC-Min") and maximum ("FC-Max") of the VaR forecasts from all 14 models for each day, are considered.
The lower tail VaR forecasts are considered here, so "FC-Min" is the most extreme of the 14 forecasts (i.e. furthest from 0) and
"FC-Max" is the least extreme. The VRates for "FC-Mean", "FC-Med", "FC-Min" and "FC-Max" series are also presented in
Tables \ref{var_fore_table} and \ref{Summ_var_fore}. Regarding these,
the "FC-Min" approach is highly conservative in each series, with few if any violations, while the "FC-Max" series produces anti-conservative
VaR forecasts that generate far too many violations. The "FC-Mean" and "FC-Median" of the 14 models produced series that generate
very competitive (sometimes the best) VRates, which is not surprising since we have approximately 50\% individual models (G-TW and RG-TWG type models) are consistently conservative and 50\% models are the opposite.

\begin{table}[!ht]
\begin{center}
\caption{\label{var_fore_table} \small 1\% VaR Forecasting VRate with different models on 7 indices and 2 assets.}\tabcolsep=10pt
\tiny
\begin{tabular}{lccccccccccc} \hline
Model           &S\&P 500       &NASDAQ          &HK              &FTSE           &DAX             &SMI            &ASX200          &IBM            &GE    \\ \hline
G-t             &\bf{1.467\%}&\bf{1.895\%}&\bf{1.652\%}&\bf{1.731\%}&1.362\%&\bf{1.617\%}&\bf{1.702\%}  &1.183\%&\fbox{0.945\%}\\
EG-t            &\bf{1.514\%}&\bf{1.611\%}&1.215\%&\bf{1.777\%}&1.408\%&\bf{1.712\%}&\bf{1.466\%}&1.183\%&\cb{1.040\%}\\
GJR-t           &\bf{1.467\%}&\bf{1.563\%}&1.263\%&\bf{1.777\%}&1.408\%&\bf{1.759\%}&\bf{1.513\%}s&1.088\%&\fbox{1.040\%}\\
Gt-HS           &1.230\%&\bf{1.563\%}&1.263\%&1.123\%&\fbox{1.127\%}&1.284\%&\fbox{0.898\%}&\fbox{1.041\%}&1.181\%\\
CARE            &1.278\%&\bf{1.563\%}&1.020\%&1.310\%&1.221\%&1.284\%&1.229\%&1.183\%&1.371\%\\
G-TW            &\fbox{0.947\%}&\cb{0.711\%}&\fbox{0.972\%}&\fbox{0.935\%}&\cb{0.704\%}&\cb{1.046\%}&0.709\%&1.088\%&0.898\%\\
RG-RV-GG        &\bf{2.130\%}&\bf{1.942\%}&\bf{2.818\%}&\bf{1.777\%}&\bf{2.300\%}&\bf{1.807\%}&\bf{1.560\%}&1.419\%&1.323\%\\
RG-RV-tG        &\bf{1.467\%}&1.326\%&\bf{1.992\%}&1.310\%&\bf{1.596\%}&1.141\%&1.229\%&0.851\%&0.803\%\\
RG-RV-TWG       &0.663\%&\fbox{0.805\%}&\bf{1.506\%}&\bf{0.608\%}&\bf{0.563\%}&0.761\%&\cb{0.851\%}&1.135\%&\fbox{0.945\%}\\
RG-RR-TWG       &\bf{0.521\%}&\bf{0.521\%}&1.166\%&\bf{0.561\%}&\bf{0.516\%}&\fbox{0.951\%}&\bf{0.567\%}&\fbox{1.041\%}&0.709\%\\
RG-ScRV-TWG     &0.615\%&0.663\%&0.875\%&0.748\%&0.610\%&0.666\%&0.662\%&1.183\%&\fbox{0.945\%}\\
RG-ScRR-TWG     &\cb{0.710\%}&0.616\%&0.826\%&0.842\%&0.657\%&0.856\%&\bf{0.567\%}&\cb{0.899\%}&0.614\%\\
RG-SubRV-TWG    &\cb{0.710\%}&\cb{0.711\%}&\fbox{0.972\%}&\cb{0.889\%}&\bf{0.563\%}&0.808\%&0.615\%&\fbox{1.041\%}&0.756\%\\
RG-SubRR-TWG    &\bf{0.521\%}&\bf{0.521\%}&\cb{1.069\%}&0.655\%&\bf{0.469\%}&\fbox{0.951\%}&0.662\%&\fbox{1.041\%}&0.662\%\\ \hline
FC-Mean       &0.899\%&0.853\%&1.166\%&0.935\%&0.704\%&1.094\%&0.709\%&0.993\%&\bf{0.473\%}\\
FC-Med    &0.947\%&0.947\%&1.166\%&0.889\%&0.751\%&1.094\%&0.757\%&0.946\%&\bf{0.567\%}\\
FC-Min       &\bf{0.189\%}&\bf{0.237\%}&\bf{0.292\%}&\bf{0.374\%}&\bf{0.235\%}&\bf{0.190\%}&\bf{0.236\%}&\bf{0.568\%}&\bf{0.142\%}\\
FC-Max &\bf{2.887\%}&\bf{2.748\%}&\bf{3.353\%}&\bf{2.993\%}&\bf{2.958\%}&\bf{2.758\%}&\bf{2.931\%}&\bf{1.845\%}&\bf{2.836\%}\\ \hline
m               &2113&2111&2058&2138&2130&2103&2115&2114&2116\\
n               &1905&1892&1890&1943&1936&1930&1871&1916&1839\\
\hline
\end{tabular}
\end{center}
\emph{Note}:\small  For individual models, box indicates the favored models based on VRate, blue shading indicates the 2nd
ranked model, whilst bold indicates the violation rate is
significantly different to 1\% by the UC test. $m$ is the out-of-sample size, and $n$ is in-sample size. RG stands for the Realized-GARCH
type models, and RC represents the Realized-CARE type models. ‘FC’ stands for forecast combination.
\end{table}

\begin{table}[!ht]
\begin{center}
\caption{\label{Summ_var_fore} \small Summary of 1\% VaR Forecast VRates, for different models on 7 indices and 2 assets.}\tabcolsep=10pt
\footnotesize
\begin{tabular}{lcc|cc|c} \hline
Model       &Mean-Overall              &Median-Overall           &Mean-Index              &Median-Index     &Mean- Assets \\ \hline
G-t& \cred{1.505\%}&\cred{1.608\%}&\cred{1.632\%}&\cred{1.609\%}&1.064\%\\
EG-t&1.437\%&1.466\%&1.530\%&1.514\%&1.111\%\\
GJR-t    &1.432\%&1.466\%&1.537\%&1.514\%&1.064\%\\
Gt-HS    &1.190\%&1.183\%&1.212\%&1.230\%&1.111\%\\
CARE          &1.274\%&1.277\%&1.273\%&1.278\%&\cred{1.277\%}\\
G-TW& \fbox{0.890\%}& \fbox{0.946\%}&\fbox{0.860\%}&\fbox{0.947\%}&\fbox{0.993\%}\\
RG-RV-GG   &\bf{1.895\%}&\bf{1.798\%}&\bf{2.045\%}&\bf{1.940\%}&\bf{1.371\%}\\
RG-RV-tG      &1.300\%&1.325\%&1.436\%&1.325\%&0.827\%\\
RG-RV-TWG        &\cb{0.869\%}&\cb{0.804\%}&\cb{0.819\%}&\cb{0.757\%}&\cb{1.040\%}\\
RG-RR-TWG        &0.726\%&0.568\%&0.684\%&0.568\%&0.875\%\\
RG-ScRV-TWG        &0.774\%&0.662\%&0.691\%&0.663\%&1.064\%\\
RG-ScRR-TWG        &0.732\%&0.710\%&0.725\%&0.710\%&0.757\%\\
RG-SubRV-TWG        &0.784\%&0.757\%&0.752\%&0.710\%&0.898\%\\
RG-SubRR-TWG        &0.726\%&0.662\%&0.691\%&0.663\%&0.851\%\\ \hline
Mean       &0.869\%&0.899\%&0.907\%&0.899\%&0.733\%\\
Median    &0.895\%&0.946\%&0.934\%&0.947\%&0.757\%\\
Min       &0.274\%&0.237\%&0.251\%&0.237\%&0.355\%\\
Max &2.811\%&2.886\%&2.946\%&2.934\%&2.340\%\\ \hline
m&2110.89&2114.00&2109.71&2113.00&2115.00\\
n&1902.44&1905.00&1909.57&1905.00&1877.50\\

\hline
\end{tabular}
\end{center}
\emph{Note}:\small  For individual models, box indicates the favoured model, blue shading indicates the 2nd ranked model, bold indicates the least favoured model,
red shading indicates the 2nd lowest ranked model, in each column. RMSE employs 1\% as the target VRate.
\end{table}

Several tests are employed to statistically assess the forecast accuracy and independence of violations from each VaR forecast model.
Table \ref{var_backtest_table} shows the number of return series (out of 9) in which each 1\% VaR forecast model is rejected for each test,
conducted at a 5\% significance level. The Re-GARCH type models are generally less likely to be rejected by the back tests
compared to other individual models, except RG-RV-GG, and the G-TW and RG-RV-tG achieved the least number of rejections (3), followed by
RG-SubRV-TWG, RG-SubRR-TWG and Gt-HS (rejected 4 times). The "FC-Mean" and "FC-Med" have very competitive results. The G-t, "FC-Min" and
"FC-Max" combinations are rejected in all 9 series, the EG-t and Re-GARCH-GG models are rejected in 8 series, respectively.

Further, in Table \ref{var_backtest_table} we add an extra column UC* showing the number of UC rejections, not counting those
when the violation rate is too conservative. For example, RG-RR-TWG of S\&P 500 was rejected with its VRate is 0.521\%, which would not be counted in
the UC* column. It is clear that most of the RG-TWG UC rejections are caused by being too conservative.

\begin{table}[!ht]
\begin{center}
\caption{\label{var_backtest_table} \small Counts of 1\% VaR  rejections with UC, CC, IND, DQ and VQR tests for different models on 7 indices and 2 assets.}\tabcolsep=10pt
\footnotesize
\begin{tabular}{lccccccccccc} \hline
mODEL&UC& UC* & CC1&IND1&DQ1&DQ4&VQR&Total\\  \hline
G-t&6&  6 &6&0&7&7&5&\bf{9}\\
EG-t&5& 5 &3&0&4&7&2&\cred{8}\\
GJR-t    &5  &5&3&0&6&5&3&7\\
Gt-HS    &1  & 1&1&0&1&3&1&\cb{4}\\
CARE          &1  &1&1&0&0&5&0&5\\
G-TW&0&  0&0&0&0&2&1&\fbox{3}\\
RG-RV-GG   &7& 7 &7&0&7&7&5&\cred{8}\\
RG-RV-tG      &3& 3 &2&0&2&1&3&\fbox{3}\\
RG-RV-TWG        &3&1  &3&1&5&3&3&6\\
RG-RR-TWG        &5& 0 &4&0&2&2&3&7\\
RG-ScRV-TWG        &0&0  &1&0&4&3&1&5\\
RG-ScRR-TWG        &1& 0 &0&0&0&2&1&\cb{4}\\
RG-SubRV-TWG        &1&0  &0&0&1&1&2&\cb{4}\\
RG-SubRR-TWG        &3& 0 &4&1&3&2&4&7\\ \hline
Mean       &1&  &1&0&0&1&0&2\\
Median    &1&  &0&0&0&2&0&2\\
Min       &9&  &8&0&8&7&9&9\\
Max &9&  &9&0&9&9&9&9\\

\hline
\end{tabular}
\end{center}
\emph{Note}:\small For individual models, box indicates the model with least number of rejections, blue shading indicates the model with 2nd least number
of rejections, bold indicates the model with the highest number of rejections, red shading indicates the model 2nd highest number of rejections.
All tests are conducted at 5\% significance level.
\end{table}

\begin{figure}[htp]
     \centering
\includegraphics[width=1.1\textwidth]{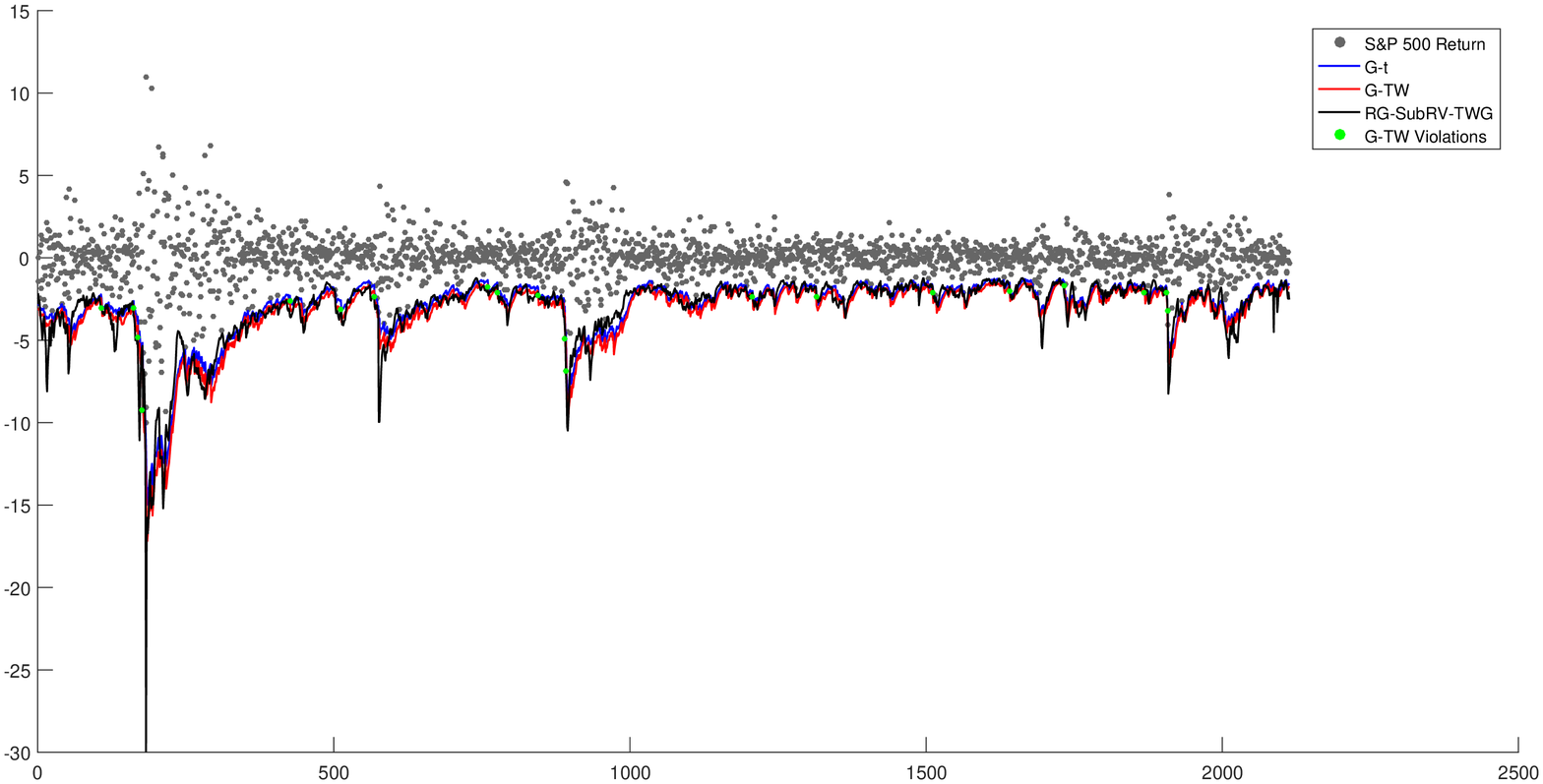}
\caption{\label{Fig_var_fore} S\&P 500 VaR Forecasts with G-t, G-TW and RG-SubRV-TWG.}
\emph{Note}:\small Returns highlighted in green are the ones exceed the VaR forecasts from G-TW.
\end{figure}

\begin{figure}[htp]
     \centering
\includegraphics[width=1.1\textwidth]{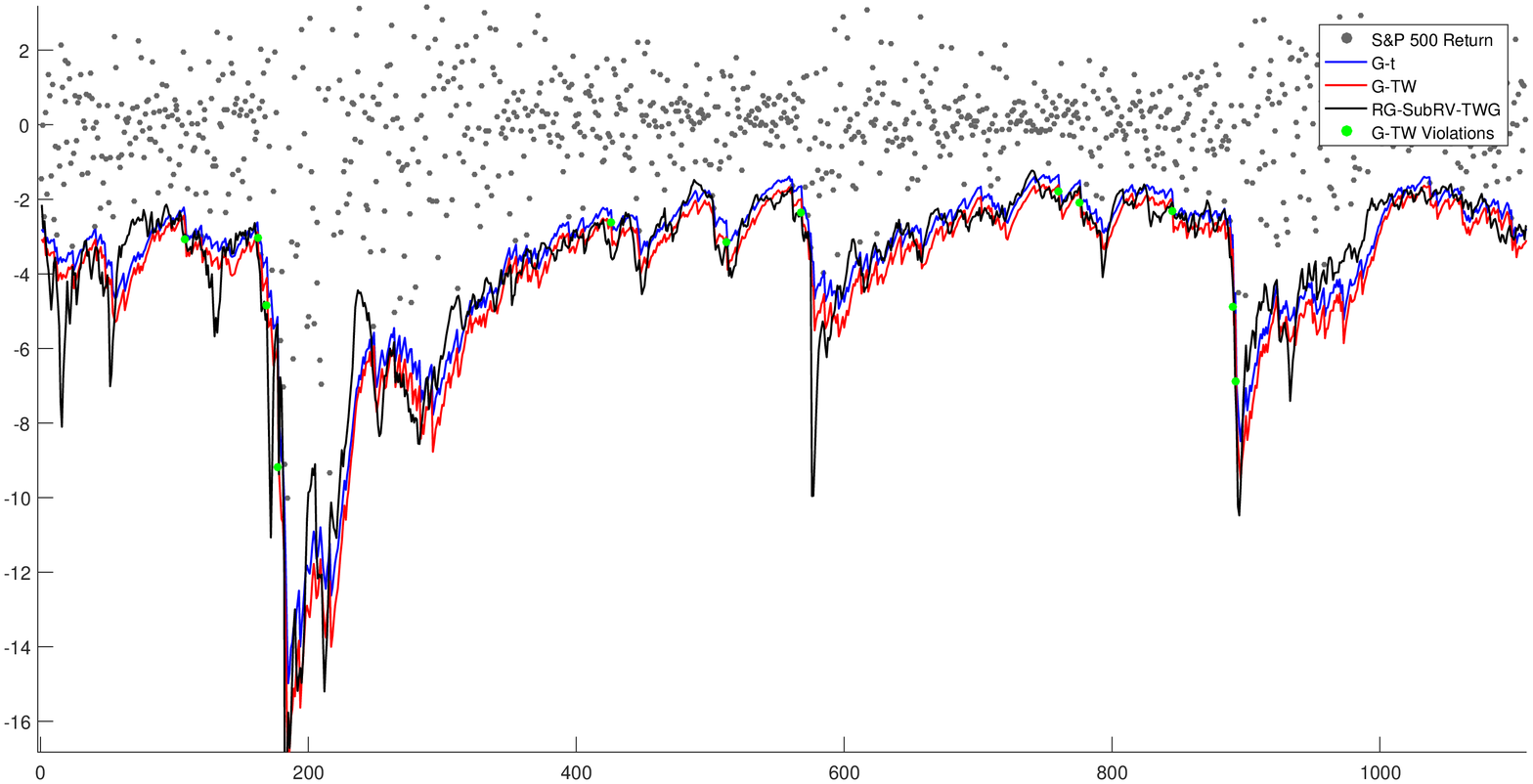}
\caption{\label{Fig_var_fore_zoom_in} S\&P 500 VaR Forecasts with G-t, G-TW and RG-SubRV-TWG (Zoomed in).}
\emph{Note}:\small Returns highlighted in green are the ones exceed the VaR forecasts from G-TW.
\end{figure}

Further, Figure \ref{Fig_var_fore} and \ref{Fig_var_fore_zoom_in} demonstrate the extra efficiency that can be gained by
employing the Re-GARCH framework with the STW distribution. Specifically, the VaR violation rates for the G-t, G-TW and RG-SubRV-TWG models are
1.467\%, 0.947\% and 0.710\% respectively, for the S\&P500 returns. These rates mean the G-t generated quite anti-conservative VaR forecasts, producing
47\% too many violations; the G-TW is more conservative and close to the perfect nominal rate; and the RG-SubRV-TWG is the most conservative model of
the three considered. Through close inspection of Figure \ref{Fig_var_fore_zoom_in}, the G-TW has an obviously more extreme (in the negative direction)
level of VaR forecasts on most days, than G-t does, but this also means the capital set aside by financial institutions to cover extreme losses,
based on such VaR forecasts, is at a higher level for the G-TW than for the G-t; this is as expected since the G-TW generates fewer violations than the
G-t in this series. However, unexpectedly, it is clearly observed that the RG-SubRV-TWG produces VaR forecasts that are often less extreme than
both the G-TW and G-t models here, meaning that lower amounts of capital are needed to protect against market risk, while simultaneously producing
a violation rate much lower than both the G-t and G-TW; the forecasts from RG-SubRV-TWG were less extreme than those from G-TW on 1546 days (73\%),
less extreme than the G-t on 915 days, in the forecast sample. This suggests a higher level of information (and cost) efficiency regarding risk
levels for the RG-SubRV-TWG model, likely coming from the increased statistical efficiency of the SubRV series over squared returns, compared to the G-t
and G-TW models, in that this model can produce VaR forecasts that have fewer violations, but are also often less extreme. Since the economic
capital is determined by financial institutions' own model and should be
directly proportional to the VaR forecast, the RG-SubRV-TWG model is able to decrease the cost capital allocation and increase the profitability of
these institutions, by freeing up part of the regulatory capital from risk coverage into investment, while still providing sufficient and more than
adequate protection against violations. The more accurate and often less extreme VaR forecasts produced by RG-SubRV-TWG are particularly
strategically important to the decision makers in the financial sector. This extra efficiency is also often observed for the RG-TWG type
models in the other markets/assets.

Further, during the GFC and other time periods with high volatility when there is a persistence of extreme returns, the
RG-SubRV-TWG VaR forecasts "recover" the fastest among the 3 models, presented through close inspection of Figure \ref{Fig_var_fore_zoom_in}, in
terms of being marginally the fastest to produce forecasts that again rejoin and follow the tail or bottom shoulder of the return data.
Traditional GARCH models tend to over-react to extreme events and to be subsequently very slow to recover, due to their oft-estimated very high
level of persistence, as discussed in Harvey and Chakravarty (2009); RG-TWG models clearly improve the performance on this aspect. Generally, the RG-SubRV model
better describes the dynamics in the volatility, compared to the traditional GARCH model, thus largely improving the responsiveness and
accuracy of the risk level forecasts, especially after high volatility periods.

\subsubsection{\normalsize VaR\&ES Joint Loss Function}
\noindent

The same set of models are employed to generate 1-step-ahead forecasts of 1\% ES during the forecast period for all 9 series.
Chen, Gerlach and Lu (2012) discuss how to treat ES forecasts as quantile forecasts in parametric models, where the quantile level
that ES falls at can be deduced exactly. Chen, Gerlach and Lu (2012) and Chen and Gerlach (2013) illustrate that across a range of
non-Gaussian distributions, when applied to financial return data, the quantile level that the 1\% ES was estimated to fall was
$\approx 0.35\%$; this value is also accurate for the Student-t and TW based models, based on their estimated parameters, for all series considered here.
Their approaches are followed to assess and test ES forecasts, by treating them as quantile forecasts and employing the UC, CC, DQ and VQR tests.

\begin{figure}[htp]
     \centering
\includegraphics[width=1.1\textwidth]{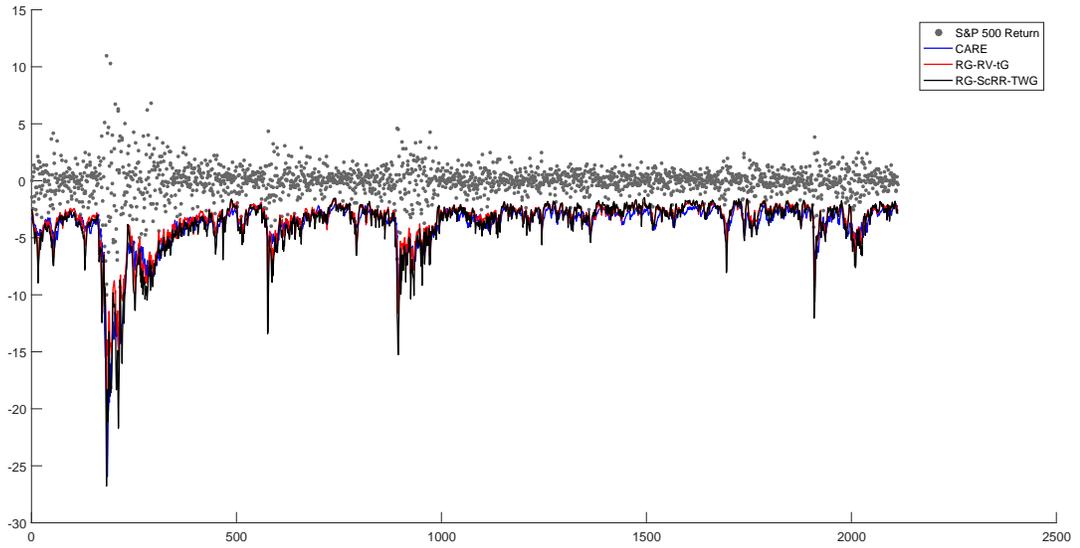}
\caption{\label{Fig_es_fore} S\&P 500 ES Forecasts with CARE, RG-RV-tG and RG-ScRR-TWG.}
\end{figure}

\begin{figure}[htp]
     \centering
\includegraphics[width=1.1\textwidth]{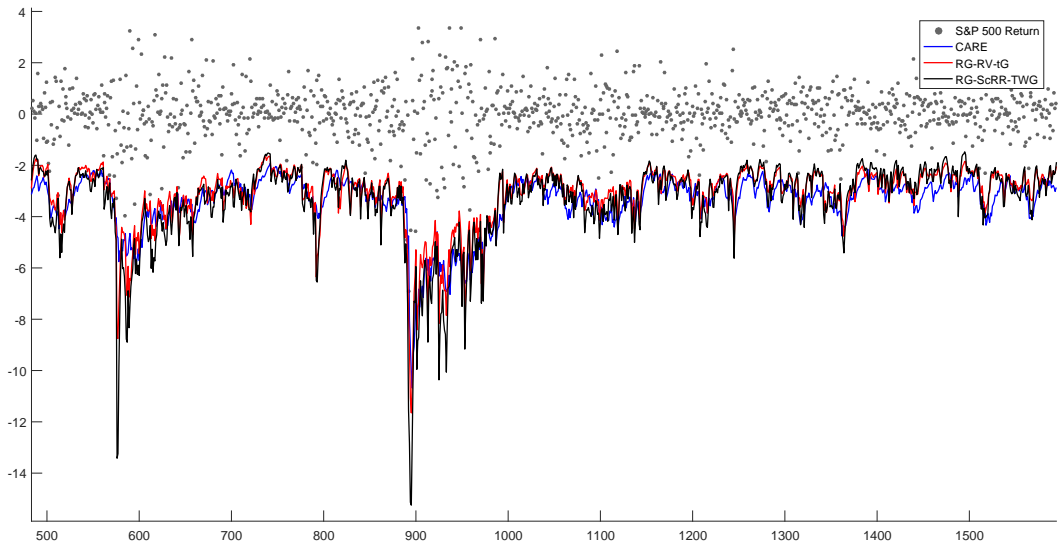}
\caption{\label{Fig_es_fore_zoom_in} S\&P 500 ES Forecasts with CARE, RG-RV-tG and RG-ScRR-TWG (Zoomed in).}
\end{figure}

First, the S\&P 500 ES forecasts with CARE, RG-RV-tG and RG-ScRR-TWG are presented in Figure \ref{Fig_es_fore} and \ref{Fig_es_fore_zoom_in}.
The ES violation rates for the 3 models are 0.284\%, 0.331\% and 0.142\% respectively. All three models generate conservative violation rates.
However, through closer inspection of the Figure \ref{Fig_es_fore_zoom_in}, the cost efficiency gains from RG-TWG models are again observed, in a
similar manner to that from the VaR forecasting study. The CARE model is reasonably conservative here, but achieves this by sacrificing efficiency: its'
ES forecasts are more extreme than the RG-ScRR-TWG model's on 1225 days (58\%). The RG-RV-tG employs the RV realized measure, which also clearly
improves the forecasting efficiency compared with the CARE from the plot. For this series the RG-RV-tG model may also be more efficient than
the RG-ScRR-TWG, generating less extreme forecasts on 74\% of days.

Cost or loss measures can be applied to assess ES forecasts, as in So and Wong (2011) who employed RMSE and MAD of the
``ES residuals'' $y_t-ES_t$, only for days when the return violates the associated VaR forecast, i.e. $y_t<VaR_t$. However, these
loss functions are not minimized by the true ES series; in fact Gneiting (2011) showed that ES is not "elicitable": i.e. there is no loss function that
is minimized by, or is strictly consistent for, the true ES series. Recently however, Fissler and Ziegel (2016) developed a family of loss
functions, that jointly assess the associated VaR and ES forecast series. This loss function family is minimized by the true VaR and ES series,
i.e. they are strictly consistent scoring functions for (VaR, ES) considered jointly. The function family form is:
\begin{eqnarray*}
S_t(y_t, VaR_t, ES_t) &=& (I_t -\alpha)G_1(VaR_t) - I_tG_1(y_t) +  G_2(ES_t)\left(ES_t-VaR_t + \frac{I_t}{\alpha}(VaR_t-y_t)\right) \\
                      &-& H(ES_t) + a(y_t) \, ,
\end{eqnarray*}
where $I_t=1$ if $y_t<VaR_t$ and 0 otherwise for $t=1,\ldots,T$, $G_1()$ is increasing, $G_2()$ is strictly increasing and strictly convex,
$G_2 = H^{'}$ and $\lim_{x\to -\infty} G_2(x) = 0$ and $a(\cdot)$ is a real-valued integrable function. Motivated by a suggestion in
Fissler and Ziegel (2016), making the choices: $G_1(x) =x$,
$G_2(x) = exp(x)$, $H(x)= exp(x)$ and  $a(y_t) = 1-\log (1-\alpha)$, which satisfy the required criteria, returns the
scoring function:
\begin{eqnarray}\label{eqveloss}
\nonumber S_t(y_t, VaR_t, ES_t) &=& (I_t -\alpha)VaR_t - I_ty_t  + \exp (ES_t) \left(ES_t-VaR_t + \frac{I_t}{\alpha}(VaR_t-y_t)\right) \\
                                &-& \exp (ES_t) + 1-\log (1-\alpha) \, ,
\end{eqnarray}
where the loss function is $S = \sum_{t-1}^T S_t$. Here, $S$ is used to informally and jointly assess and compare the VaR and ES forecasts
from all models.

Table \ref{veloss} shows the loss function values $S$, calculated using equation (\ref{eqveloss}), which jointly assess the
accuracy of each model's VaR and ES series, during the forecast period for each market. On this measure, the RG-TWG models using SubRV and
ScRR do best overall, having lower loss than most other models in most series and
being consistently ranked lower on that measure. The EGARCH and GARCH, with Student-t errors, and CARE models models consistently rank lowest
among the individual models, only trailed by the forecast combination method "FC-Max"; for models with TW errors, the G-TW is consistently rank
lowest; the realized GARCH models consistently rank the highest, with lowest loss. These rankings are consistent with the findings illustrated in
figures \ref{Fig_var_fore} and \ref{Fig_var_fore_zoom_in}. Although the G-TW generated the VRate closest to the nominal 1\% level, it is
excessively conservative in the forecasting period, in terms of capital allocation, thus sacrificing efficiency, requiring institutions to deploy
higher economic capital to cover potential extreme losses. Apparently, the proposed RG-TW type models demonstrate an advantage regarding this aspect.
To conclude, the RG-TW models have lower loss in joint VaR and ES prediction and are higher ranked than other models in most markets/assets.
These models, together with the "FC-Med" and "FC-Mean", consistently outperform all the other models.

\begin{table}[!ht]
\begin{center}
\caption{\label{veloss} \small VaR and ES joint loss function values, using equation (\ref{eqveloss}), across the markets; $\alpha=0.01$.}\tabcolsep=10pt
\tiny
\begin{tabular}{lccccccccccccc} \hline
Model           &S\&P 500         &NASDAQ         &HK             &FTSE           &DAX            &SMI            &ASX200         &IBM            &GE    \\ \hline
G-t             &\cred{2119.20}&2157.13&2135.83&2156.41&2226.67&2153.91&2082.44&2270.94&2229.71\\
EG-t            &\bf{2136.42}&\cred{2167.79}&2121.78&\bf{2187.08}&\bf{2239.32}&\cred{2161.34}&\cred{2095.26}&2285.78&\cred{2230.77}\\
GJR-t           &2099.80&2140.86&2120.74&2156.49&\cred{2238.97}&\bf{2175.85}&2077.59&\cred{2287.56}&2230.25\\
Gt-HS           &2109.76&2148.21&2128.70&2139.18&2219.45&2123.79&2075.06&2257.37&2228.86\\
CARE            &2116.00&\bf{2182.52}&2117.80&\cred{2156.71}&2202.48&2137.79&\bf{2136.67}&\cb{2232.60}&\bf{2321.08}\\
G-TW         &2096.78&2136.81&2123.45&2136.17&2205.71&2118.14&2077.59&2275.71&2228.82\\
RG-RV-GG        &2093.47&2146.03&\bf{2217.36}&2134.75&2214.21&2138.85&2067.2&\bf{2319.63}&2202.88\\
RG-RV-tG        &2070.69&2128.77&2146.84&\fbox{2116.82}&2185.87&2107.73&\fbox{2051.68}&\fbox{2230.69}&2204.20\\
RG-RV-TWG       &2083.04&2130.08&2141.81&2129.77&2186.06&2142.85&\cb{2064.92}&2248.23&2267.11\\
RG-RR-TWG       &2062.02&2117.91&2142.81&2127.45&2184.12&2093.06&2084.42&2266.91&\fbox{2194.86}\\
RG-ScRV-TWG     &2075.51&2130.26&\cred{2148.66}&2125.47&2188.48&2134.14&2066.80&2248.52&2265.06\\
RG-ScRR-TWG     &2062.83&\fbox{2112.49}&\cb{2115.28}&2121.91&\fbox{2183.07}&\fbox{2090.86}&2087.29&2265.25&2208.61\\
RG-SubRV-TWG    &\fbox{2055.33}&\cb{2112.64}&\fbox{2112.16}&\cb{2120.18}&\cb{2183.78}&\cb{2091.06}&2065.97&2262.32&2209.53\\
RG-SubRR-TWG    &\cb{2055.34}&2117.03&2130.03&2124.39&2183.94&2092.61&2073.19&2264.93&\cb{2202.08}\\ \hline
FC-Mean       &2068.23&2117.87&2111.74&2113.98&2180.43&2097.4&2055.38&2247.15&2219.66\\
FC-Med        &2066.54&2115.95&2107.83&2115.24&2180.75&2098.30&2054.33&2250.18&2214.41\\
FC-Min        &2111.73&2153.94&2149.29&2141.95&2204.87&2143.43&2103.11&2237.43&2258.02\\
FC-Max        &2174.45&2224.42&2253.19&2258.19&2297.20&2235.48&2190.91&2353.91&2297.93\\ \hline
\end{tabular}
\end{center}
\emph{Note}:\small  For individual models, box indicates the favoured model, blue shading indicates the 2nd ranked model, bold indicates the least
favoured model, red shading indicates the 2nd lowest ranked model, in each column.
\end{table}

\begin{table}[!ht]
\begin{center}
\caption{\label{veloss)summary} \small VaR and ES joint loss function values summary; $\alpha=0.01$.}\tabcolsep=10pt
\begin{tabular}{lccccccccccccc} \hline
Model           & Mean loss          & Mean rank  \\ \hline
G-t&2170.25&\cred{13.88}\\
EG-t&\bf{2180.62}&\bf{15.00}\\
GJR-t    &2169.79&13.00\\
Gt-HS    &2158.93&10.88\\
CARE          &2178.18&11.63\\
G-TW&2155.46&10.50\\
RG-RV-GG   &\cred{2170.49}&12.25\\
RG-RV-tG      &\cb{2138.14}&5.88\\
RG-RV-TWG        &2154.87&8.50\\
RG-RR-TWG        &2141.51&8.13\\
RG-ScRV-TWG        &2153.65&8.75\\
RG-ScRR-TWG        &2138.62&\cb{5.38}\\
RG-SubRV-TWG        &\fbox{2134.77}&\fbox{3.75}\\
RG-SubRR-TWG        &2138.17&6.00\\ \hline
FC-Mean       &2134.65&3.38\\
FC-Med    &2133.73&3.50\\
FC-Min       &2167.08&12.63\\
FC-Max &2253.96&18.00\\ \hline
\end{tabular}
\end{center}
\emph{Note}:\small  For individual models, boxes indicate the favoured model, blue shading indicates the 2nd ranked model, bold indicates the least
favoured model, red shading indicates the 2nd lowest ranked model, in each column. "Mean rank" is the average rank across the 7 markets and 2 assets
for the loss function, over the 19 models: lower is better.
\end{table}

\subsubsection{\normalsize Model confidence set}
\noindent
The model confidence set (MCS), introduced by Hansen, Lunde and Nason (2011), is a method to statistically compare a
group of forecast models via a loss function. MCS is applied here to further compare among the 19 (VaR, ES) forecasting models.
A MCS is a set of models that is constructed such that it will contain the
best model with a given level of confidence, which was selected as 90\% in our paper; Matlab code for MCS testing was
downloaded from "www.kevinsheppard.com/MFE\_Toolbox". We adapted the code to incorporate the VaR and ES joint loss function values
(Equation, \ref{eqveloss}) as the loss function during the MCS calculation. Two methods (R and SQ) to calculate the test statistics
are employed to the MCS selection process.

Table \ref{mcs_r} and \ref{mcs_sq} present the 90\% MCS using the R and SQ methods, respectively. Column "Total" counts the
total number of times that a model is included in the 90\% MCS across the 9 return series. Based on this column, boxes indicate the
favoured model, and blue shading indicates the 2nd ranked model for each market. Bold indicates the least favoured and red shading
indicates the 2nd lowest ranked model for each market.

Via the R method, RG-SubRV-TWG has the best performance and was included in the MCS for 8 markets and assets, followed by RG-TWG with RR, ScRR and
RG-RV-tG (included 7times in the MCS in 9 series). "G-t" is only included in the 90\% MCS once. Via the SQ method, the proposed
RG-TWG models are still favoured. The 90\% MCS includes RG-SubRV-TWG, RG-SubRR-TWG and RG-RV-tG in all 9 series,
followed by RG-ScRR-TWG (8 times). For either R or SQ methods, "FC-Mean" and "FC-Med" still have quite competitive performances.

\begin{table}[!ht]
\begin{center}
\caption{\label{mcs_r} \small 90\% model confidence set with R method across the markets and assets.}\tabcolsep=10pt
\tiny
\begin{tabular}{lccccccccccccc} \hline
Model    &S\&P 500  &NASDAQ &HK  &FTSE  &DAX  &SMI   &ASX200 &IBM &GE  &Total  \\ \hline
G-t&0&0&0&0&0&0&0&1&0&\bf{1}\\
EG-t&0&0&1&0&0&1&0&1&0&3\\
GJR-t    &0&1&1&1&0&0&1&1&0&5\\
Gt-HS    &0&0&1&1&0&1&0&1&0&4\\
CARE          &0&0&1&0&0&0&0&1&0&\cred{2}\\
G-TW&0&0&1&0&0&1&0&1&0&3\\
RG-RV-GG   &1&0&0&1&0&1&1&1&1&6\\
RG-RV-tG      &1&1&0&1&1&1&1&1&0&\cb{7}\\
RG-RV-TWG        &0&0&0&1&1&0&1&1&0&4\\
RG-RR-TWG        &1&1&0&1&1&1&0&1&1&\cb{7}\\
RG-ScRV-TWG        &1&0&0&1&1&0&1&1&0&5\\
RG-ScRR-TWG        &1&1&1&1&1&1&0&1&0&\cb{7}\\
RG-SubRV-TWG        &1&1&1&1&1&1&1&1&0&\fbox{8}\\
RG-SubRR-TWG        &1&1&0&1&1&1&0&1&0&6\\  \hline
Mean       &1&1&1&1&1&1&1&1&0&8\\
Median    &1&1&1&1&1&1&1&1&0&8\\
Min       &0&0&0&0&0&0&0&1&0&1\\
Max &0&0&0&0&0&0&0&0&0&0\\  \hline
\end{tabular}
\end{center}
\emph{Note}:\small For individual models, boxes indicate the favoured model, blue shading indicates the 2nd ranked model, bold indicates the least
favoured model, red shading indicates the 2nd lowest ranked model, based on total number of included in the MCS across the 7 markets and 2 assets,
higher is better.
\end{table}

\begin{table}[!ht]
\begin{center}
\caption{\label{mcs_sq} \small 90\% model confidence set with SQ method across the markets and assets.}\tabcolsep=10pt
\tiny
\begin{tabular}{lccccccccccccc} \hline
Model    &S\&P 500  &NASDAQ &HK  &FTSE  &DAX  &SMI   &ASX200 &IBM &GE  &Total  \\ \hline
G-t&0&0&1&1&1&0&1&1&0&\cred{5}\\
EG-t&0&0&1&1&0&1&1&1&0&\cred{5}\\
GJR-t    &0&1&1&1&0&0&1&1&0&\cred{5}\\
Gt-HS    &0&0&1&1&0&1&1&1&1&6\\
CARE          &0&0&1&1&1&0&0&1&0&\bf{4}\\
G-TW&0&0&1&1&0&1&1&1&0&\cred{5} \\
RG-RV-GG   &0&0&0&1&1&1&1&1&1&6\\
RG-RV-tG      &1&1&1&1&1&1&1&1&1&\fbox{9}\\
RG-RV-TWG        &0&0&1&1&1&0&1&1&0&\cred{5}\\
RG-RR-TWG        &1&1&0&1&1&1&0&1&1&7\\
RG-ScRV-TWG        &0&0&0&1&1&0&1&1&0&\bf{4}\\
RG-ScRR-TWG        &1&1&1&1&1&1&0&1&1&\cb{8}\\
RG-SubRV-TWG        &1&1&1&1&1&1&1&1&1&\fbox{9}\\
RG-SubRR-TWG        &1&1&1&1&1&1&1&1&1&\fbox{9}\\  \hline
Mean       &1&1&1&1&1&1&1&1&0&8\\
Median    &1&1&1&1&1&1&1&1&0&8\\
Min       &0&0&0&1&1&0&0&1&0&3\\
Max &0&0&0&0&0&0&0&0&0&0\\  \hline
\end{tabular}
\end{center}
\emph{Note}:\small For individual models, boxes indicate the favoured model, blue shading indicates the 2nd ranked model, bold indicates the least
favoured model, red shading indicates the 2nd lowest ranked model, based on total number of included in the MCS across the 7 markets and 2 assets, higher
is better.
\end{table}

\section{Conclusion}\label{conclusion_section}
\noindent
In this paper, the Realized-GARCH is extended through incorporating the two-sided Weibull distribution to estimate and forecast financial tail risk.
In addition, the scaled and sub-sampled realized measures have been incorporated into the proposed Re-GARCH-TWG framework, aiming to further improve
the out-of-sample forecasting of the proposed model. The proposed RG-TWG type models generated consistently adequately sufficient
risk coverage and relatively accurate, conservative violation rates, compared to competing models including Re-GARCH models employing realized volatility,
traditional GARCH, CARE models and a GARCH with two-sided Weibull distribution. Forecast combinations methods employing the mean and median of the
forecasts also produce very competitive tail risk forecasting results, which is related to the fact that the proposed RG-TWG type models are
always conservative. Regarding back testing of VaR forecasts, the RG-TWG type models are also generally less likely to be rejected than their
counterparts. With respect to the VaR and ES joint loss function values, RG-TWG model's VaR and ES forecasts consistently had lower loss than all
other models considered, especially those employing sub-sampled RV and scaled RR. The combined series "FC-Mean" and "FC-Med" are also highly
competitive regarding this loss function. Further, the model confidence set results also favour the proposed RG-TWG framework, especially the
ones incorporating SubRV and SubRR, as well as the standard Re-GARCH with Student-t errors. In addition to being more conservative and accurate
via minimising loss, the RG-TWG models is also more efficient, through generating less extreme tail risk forecasts and regularly allowing smaller
amounts of capital allocation without being anti-conservative or significantly inaccurate. To conclude, the RG-TWG type
models with sub-sampled RV and scaled RR should be considered for financial applications when
forecasting tail risk, and should allow financial institutions to more accurately and efficiently allocate capital under the Basel Capital Accord,
to protect their investments from extreme market movements. This work could be extended by considering more distributions for the return equation
and alternative distributions for the measurement equation and by using alternative frequencies of observation for the realized measures.

\clearpage
\section*{References}
\addcontentsline{toc}{section}{References}
\begin{description}

\item Andersen, T. G. and Bollerslev, T. (1998). Answering the skeptics: Yes, standard volatility models do provide accurate
forecasts. \emph{International economic review}, 885-905.

\item Andersen, T. G., Bollerslev, T., Diebold, F. X. and Labys, P. (2003). Modeling and forecasting realized volatility.
    \emph{Econometrica}, 71(2), 579-625.

\item Artzner, P., Delbaen, F., Eber, J.M., and Heath, D. (1997). Thinking coherently.  \emph{Risk}, 10, 68-71.

\item Artzener, P., Delbaen, F., Eber, J.M., and Heath, D. (1999). Coherent measures of risk.  \emph{Mathematical Finance}, 9, 203-228.

\item Black, F. (1976). Studies in stock price volatility changes. \emph{In: American Statistical Association Proceedings of the Business and Economic Statistics Section}. 177–181.

\item Bollerslev, T. (1986). Generalized Autoregressive Conditional Heteroskedasticity. \emph{Journal of Econometrics}, 31, 307-327.

\item Bollerslev, T. (1987). A conditionally heteroskedastic time series model for speculative prices and rates of return. \emph{The review of economics and statistics}, 542-547.

\item Chang, C. L., Jim{\'e}nez-Mart{\'i}n, J. {\'A}., McAleer, M., and P{\'e}rez-Amaral, T. (2011). Risk management of risk under the
    Basel Accord: Forecasting value-at-risk of VIX futures.  \emph{Managerial Finance}, 37, 1088-1106.

\item Chen, C.W.S., Gerlach, R. and So, M.K.P. (2008). Comparison of nonnested asymmetric heteroskedastic models.
  \emph{Computational Statistics and Data Analysis}, 51(4), 2164-2178.

\item Chen, Q., Gerlach, R. and Lu, Z. (2012). Bayesian Value-at-Risk and expected shortfall
forecasting via the asymmetric Laplace distribution. \emph{Computational Statistics and Data
Analysis}, 56, 3498-3516.

\item Chen, Q. and Gerlach, R. (2013). The two-sided Weibull distribution and forecasting financial tail risk. \emph{International Journal of Forecasting}, 29(4),527-540.

\item Chen, W., Peters, G., Gerlach, R. and Sisson, S. (2017). Dynamic Quantile Function Models. arXiv:1707.02587.

\item Christensen, K. and Podolskij, M. (2007). Realized range-based estimation of integrated variance. \emph{Journal of Econometrics},
141(2), 323-349.

\item Christoffersen, P. (1998). Evaluating interval forecasts. \emph{International Economic Review}, 39, 841-862.

\item Contino, C. and Gerlach, R. (2017). Bayesian tail-risk forecasting using realized GARCH. \emph{Applied Stochastic Models in Business and Industry}, 33:2, 213-36.

\item Engle, R. F. (1982), Autoregressive Conditional Heteroskedasticity with Estimates of the Variance of United Kingdom
Inflations. \emph{Econometrica}, 50, 987-1007.

\item Engle, R. F. and Manganelli, S. (2004). CAViaR: Conditional Autoregressive Value at Risk
by Regression Quantiles. \emph{Journal of Business and Economic Statistics}, 22, 367-381.

\item Engle, R.F. and Russell, J. (1998). Autoregressive conditional duration: A new model for irregulatory spaced transaction data. \emph{Econometrica}, 66, 1127-1162.


\item Fissler, T. and Ziegel, J. F. (2016). Higher order elicibility and Osband's principle. \emph{Annals of Statistics}, in press.

\item Gaglianone, W. P., Lima, L. R., Linton, O. and Smith, D. R. (2011). Evaluating Value-
at-Risk models via quantile regression. \emph{Journal of Business and Economic Statistics},
29, 150-160.




\item Gerlach, R., Walpole, D. and Wang, C. (2016). Semi-parametric Bayesian Tail Risk Forecasting Incorporating Realized Measures of Volatility,
\emph{Quantitative Finance}, 17:2, 199-215 .

\item Gerlach, R. and Wang, C. (2016).  Forecasting risk via realized GARCH, incorporating the realized range.
\emph{Quantitative Finance}, 16:4, 501-511.

\item Glosten, L.R., Jagannathan, R. and Runkle, D.E. (1993). On the relation between the expected value and the volatility of the nominal excess return on stocks. \emph{The journal of finance}, 48(5), pp.1779-1801.

\item Gneiting, T (2011). Making and evaluating point forecasts. \emph{Journal of the American Statistical Association}, 106,
494, 746-762.

\item Hansen, B. E. (1994). Autoregressive conditional density estimation. \emph{International Economic Review}, 35, 705-730.


\item Hansen, P. R., Huang, Z. and Shek, H. H. (2011). Realized GARCH: a joint model for returns and realized measures of volatility.
\emph{Journal of Applied Econometrics}, 27(6), 877-906.

\item Hansen, P.R., Lunde, A. and Nason, J.M. 2011. The model confidence set.  \emph{Econometrica}, 79(2), 453-497.

\item Harvey, A.C. and T. Chakravarty (2009). Beta-t-(E)GARCH. Cambridge Working Papers in Economics 0840, Faculty of Economics, University of Cambridge, Cambridge.


\item Kupiec, P. H. (1995). Techniques for Verifying the Accuracy of Risk Measurement Models. \emph{The Journal of Derivatives}, 3, 73-84.

\item Malevergne, Y. and Sornette, D. (2004). VaR-efficient portfolios for a class of super and sub-exponentially decaying assets return distributions. \emph{Quantitative Finance}, 4, 17-36.

\item Martens, M. and van Dijk, D. (2007). Measuring volatility with the realized range. \emph{Journal of Econometrics}, 138(1), 181-207.

\item McAleer, M., Jim{\'e}nez-Mart{\'i}n, J. {\'A}. and P{\'e}rez-Amaral, T. (2013), GFC-robust risk management strategies under the
    Basel Accord,  \emph{International Review of Economics and Finance}, 27 , pp. 97-111.

\item Mittnik, M. and Rachev, S.T. (1989). Stable distributions for asset returns. \emph{Applied Mathematics Letters}, 2, 301-304.


\item Metropolis, N., Rosenbluth, A. W., Rosenbluth, M. N., Teller, A. H., and Teller, E. (1953). Equation of State Calculations by Fast
Computing Machines. \emph{J. Chem. Phys}, 21, 1087-1092.

\item Nelson, D. B. (1991). Conditional Heteroskedasticity in Asset Returns: A New Approach. \emph{Econometrica}, 59, 347-370.

%


\item Roberts, G. O., Gelman, A. and Gilks, W. R. (1997). Weak convergence and optimal scaling of random walk Metropolis algorithms.
    \emph{The annals of applied probability}, 7(1), 110-120.


\item So, M.K.P and Wong, C.M. (2011). Estimation of multiple period expected shortfall and median shortfall for risk management.
 \emph{Quantitative Finance}, 1-16.

\item Taylor, J. (2008). Estimating Value at Risk and Expected Shortfall Using Expectiles. \emph{Journal of Financial Econometrics}, 6,
    231-252.

\item Watanabe, T. (2012). Quantile Forecasts of Financial Returns Using Realized GARCH Models. \emph{Japanese Economic Review}, 63(1),
    68-80.

\item Weibull, W. (1951). A statistical distribution function of wide applicability. \emph{Journal of Applied Mechanics -Trans. ASME}, 18, 293-297.


\item Zhang, L., Mykland, P. A., and A\"{i}t-Sahalia, Y. (2005). A tale of two time scales.  \emph{Journal of the American Statistical
    Association}, 100(472).

\end{description}

\end{document}